\documentclass{article}

\usepackage[dblblindworkshop, final]{neurips_2026}
\workshoptitle{AI4DD: AI for Drug Discovery -- Bridging the Translation Gap}

\usepackage[utf8]{inputenc}
\usepackage[T1]{fontenc}
\usepackage{hyperref}
\usepackage{url}
\usepackage{booktabs}
\usepackage{amsfonts}
\usepackage{amsmath}
\usepackage{nicefrac}
\usepackage{microtype}
\usepackage{xcolor}
\usepackage{graphicx}
\usepackage{wrapfig} 

\usepackage{enumitem}

\title{The Dose Makes the Poison: Exposure-Aware Evaluation of Generative Drug Design}

\author{%
  Maria Franz \\
  International Computer Science Institute \\
  Berkeley, CA, USA \\
  \texttt{mfranz@icsi.berkeley.edu} \\
  \And
  N. Benjamin Erichson \\
  International Computer Science Institute \\
  Berkeley, CA, USA \\
  \texttt{erichson@icsi.berkeley.edu} \\
}

\begin{document}

\maketitle

\begin{abstract}
Generative models are increasingly used to propose molecules for drug discovery. They can produce far more candidates than can be experimentally tested, making computational selection essential. Most selection methods rely on predicted molecular properties, such as potency and toxicity, but overlook how these properties translate into drug exposure at a given dose. As a result, molecules that appear promising on paper may require impractical doses or raise safety concerns. We investigate this gap by using physiologically based pharmacokinetic (PBPK) simulations to re-evaluate molecules selected from three PARP1-targeted generative pipelines. Of 60 shortlisted molecules, seven fail to achieve sufficient target exposure, while $28\%$ of the remaining candidates show predicted safety risks at their required dose. Moreover, $85\%$ of candidates with an achievable dose are outperformed by molecules overlooked during static selection. These findings suggest that accounting for dose and exposure can substantially change which generated molecules are considered promising for drug development.
\end{abstract}

\section{Introduction}
\label{sec:introduction}

A challenge in computational drug discovery is deciding which molecules are worth pursuing. Generative models can explore large chemical spaces and propose thousands of candidates, but only a small fraction can be experimentally tested. Selecting promising molecules therefore requires computational methods to assess their properties before experimental validation. Most current workflows rely on predictions of potency, toxicity, and drug-like properties. Such predictions are widely used to benchmark generative models \citep{gao2022sample,brown2019guacamol,polykovskiy2020moses,garciaortegon2022dockstring}, guide molecular generation \citep{loeffler2024reinvent4,lee2023mood}, and filter or rank candidates using multi-parameter scoring functions \citep{wager2010cnsmpo}.

While useful, these predictions do not tell us whether a molecule will perform well at the dose needed for treatment. Both efficacy and toxicity depend on drug exposure, which is determined by the interaction of several properties, including potency, absorption, clearance, and protein binding \citep{maurer2020dose}. For example, a molecule with high predicted potency may still require a large dose if it is rapidly cleared or poorly absorbed. Similarly, a molecule with weak predicted off-target activity may become a safety concern if the dose required for target engagement leads to high systemic exposure. Evaluating molecular properties independently can therefore miss important differences between candidates.
Physiologically based pharmacokinetic (PBPK) models offer a way to account for these effects by simulating how drug concentrations change over time following administration. PBPK modeling is widely used in drug development, including for target pharmacology and dose prediction \citep{chen2021mtpa,geci2024systematic,naga2022htpbpk}. Recent studies have also explored incorporating predicted pharmacokinetic properties or PBPK-derived exposure into molecular generation \citep{maagdenberg2025pkqsp,ramey2022optimal}, as well as improving potency while preserving favorable bioavailability and toxicity properties \citep{horne2024generative}. In this work, we take a different approach. Rather than modifying the generative objective, we use PBPK simulations to examine whether molecules selected by conventional property-based screening remain promising once dose and exposure are considered.

We investigate this question using three PARP1-targeted generative pipelines: \emph{MOOD}, a docking-guided diffusion model \citep{lee2023mood}; \emph{REINVENT4-MPO}, a recurrent neural network fine-tuned through reinforcement learning to optimize potency and ADME-related properties; and \emph{Mol2Mol}, a seed-based approach that generates analogues of existing molecules. Both REINVENT4-MPO and Mol2Mol are implemented using the REINVENT4 framework \citep{loeffler2024reinvent4}.
We treat these pipelines as separate case studies rather than a controlled comparison. For each pipeline, we apply the same property-based filters and static ranking procedure, then re-evaluate the shortlisted molecules using PBPK simulations. We estimate the dose needed to achieve sufficient target exposure and assess the associated safety margins for two liabilities relevant to PARP inhibitors.

Our main contributions are:

\begin{itemize}[leftmargin=*]
    \item We introduce an exposure-aware evaluation of generated molecules that uses PBPK simulations to connect predicted molecular properties with dose requirements and safety margins.

    \item We show that favorable molecular property predictions do not always translate into favorable exposure. Of 60 shortlisted molecules, seven cannot achieve sufficient target exposure under our dosing constraints. Of those that can, 28\% show predicted safety risks at their required dose.
    
    \item We demonstrate that accounting for dose and exposure can change candidate selection. Among the 53 shortlisted molecules with an achievable dose, 45 are outperformed in dose requirement and both safety margins by at least one molecule overlooked by the static ranking.
\end{itemize}

\section{Method: Exposure-Aware Evaluation of Generated Molecules}
\label{sec:methods}

We evaluate generated molecules based on the dose needed to achieve sufficient target exposure and the associated safety margins. For each candidate, we combine predicted molecular properties with physiologically based pharmacokinetic (PBPK) simulations to estimate drug exposure following oral administration \citep{chen2021mtpa}. We then determine the dose required for target engagement and assess safety at that dose, accounting for uncertainty in both the molecular predictions and simulated exposure. This allows us to compare candidates based on their predicted behavior at a relevant dose.

\textbf{From Molecular Properties to Exposure.}
For each candidate, we use predicted physicochemical and pharmacokinetic properties to parameterize an oral PBPK model. The model simulates drug concentrations over time at a given dose, accounting for the combined effects of properties such as absorption, clearance, and protein binding \citep{naga2022htpbpk}. These properties influence not only how much of a drug reaches circulation, but also how long it remains there. Two molecules with similar predicted potency may therefore produce very different concentration profiles at the same dose. By simulating these profiles, we can assess whether a candidate achieves sufficient target exposure and how close its concentrations come to predicted off-target safety thresholds.

\textbf{Dose Required for Target Engagement.}
We first determine whether a candidate can achieve sufficient target exposure and, if so, at what dose. As our target threshold, we use the predicted half-maximal inhibitory concentration ($IC50_{\mathrm{target}}$), which is the concentration required to inhibit 50\% of the target's activity in an assay. We then search for the minimum oral dose at which the simulated free (unbound) drug concentration remains above this threshold during the dosing interval:
\begin{equation}
d^{*}
=
\min_{d}
\left\{
d :
C_{\mathrm{trough,free}}(d)
\geq
IC50_{\mathrm{target}}
\right\}.
\end{equation}
Here, $C_{\mathrm{trough,free}}(d)$ is the lowest free drug concentration between doses after repeated administration has reached steady state. We refer to $d^{*}$ as the minimum effective dose (MED). This is a model-based estimate, not a clinically established effective dose. Using the trough concentration provides a conservative criterion for maintaining the predicted target exposure throughout the dosing interval. Importantly, MED depends on both potency and pharmacokinetics: even a highly potent molecule may require a large dose if it is poorly absorbed or rapidly eliminated \citep{maurer2020dose}.
We determine MED iteratively using oral PBPK simulations, considering once- or twice-daily administration. For each candidate, we simulate repeated dosing and adjust the dose until the target concentration is reached. If no dose achieves the required exposure within our search constraints, we report that no achievable MED was found. Details of the dose-search procedure are provided in Appendix~\ref{app:med}.

\textbf{Safety at the Required Dose.}
A candidate that reaches its target concentration is not necessarily safe. The dose needed for target engagement may also produce concentrations high enough to inhibit other proteins, potentially causing adverse effects. We therefore evaluate safety at each candidate's own MED, rather than comparing off-target potency predictions in isolation.
For safety endpoint $j$, we define the exposure-adjusted safety margin as
\begin{equation}
M_j
=
\frac{IC50_j}
     {C_j(d^{*})},
\end{equation}
where $IC50_j$ is the predicted concentration required to inhibit the off-target and $C_j(d^{*})$ is the relevant simulated drug concentration at the MED. A margin greater than one means that the predicted exposure is below the off-target inhibition threshold, while a margin below one means that it exceeds this threshold. Larger margins are therefore preferable. Two molecules with similar off-target potency can have different safety margins if they require different exposures to engage their intended target.

We consider two safety endpoints relevant to PARP inhibitors. The first is hERG, a cardiac potassium channel whose inhibition is associated with arrhythmia risk. We evaluate hERG using peak drug concentration, $C_{\max}$, because cardiac safety assessments commonly compare peak unbound plasma exposure with hERG inhibitory potency \citep{redfern2003relationships}. The second is PARP2, a related enzyme whose inhibition can contribute to hematological toxicity. For PARP2, we use steady-state trough concentration, $C_{\mathrm{trough}}$, to assess sustained off-target exposure. These endpoints capture different aspects of the concentration--time profile: peak exposure and the concentration maintained between doses.
Both predicted off-target potency and simulated exposure are uncertain, so safety margins based on central estimates alone may be misleading. We therefore classify candidates into three risk categories. A candidate is \emph{RED} if its central predicted exposure exceeds the off-target potency threshold ($M_j < 1$), \emph{ORANGE} if the central exposure is below the threshold but the uncertainty intervals overlap, and \emph{GREEN} if the intervals remain separated. This distinguishes candidates with predicted safety concerns from those whose apparent safety remains uncertain. These categories reflect model-based risk, not experimentally established toxicity or safety. Uncertainty estimates and their propagation are described in Appendix~\ref{app:pbpk-error}.

\section{Experimental Setup}
\label{sec:experimental-setup}

We investigate whether accounting for dose and exposure changes which molecules are selected by a conventional property-based screening workflow. For each of three generative pipelines, we generate a pool of candidates, apply drug-likeness and activity filters, and rank the remaining molecules using a static multi-parameter score. We then select 20 molecules per pipeline for the exposure-aware evaluation described in Section~\ref{sec:methods}. Applying the same procedure to each pool allows us to compare static rankings with exposure-aware outcomes in terms of candidate selection.

\textbf{Generative pipelines.}
We study three PARP1-targeted pipelines that differ in how they generate and optimize molecules. \emph{MOOD} \citep{lee2023mood} is a diffusion model guided by a predicted PARP1 docking score, without explicitly considering absorption, distribution, metabolism, or excretion (ADME) during generation. \emph{REINVENT4-MPO} uses a recurrent neural network fine-tuned through reinforcement learning to optimize predicted PARP1 potency alongside physicochemical and ADME-related properties. \emph{Mol2Mol} generates analogues of existing molecules, starting from 20 approved PARP1 inhibitors. Both REINVENT4-MPO and Mol2Mol are implemented using the REINVENT4 framework \citep{loeffler2024reinvent4}.
We generate 10{,}000 molecules each from MOOD and REINVENT4-MPO, and 1{,}000 from Mol2Mol (50 analogues per seed). Because the pipelines differ in architecture, training data, and starting chemical space, we treat them as independent case studies rather than a controlled comparison of generative methods. This allows us to examine the same selection question across different generative settings without attributing differences in pharmacokinetic behavior to any single modeling choice. Implementation details are provided in Appendix~\ref{app:generation}.

\textbf{Filtering and static candidate selection.}
All generated molecules undergo the same filtering and ranking procedure. We first apply hard filters for drug-likeness, synthetic accessibility, physicochemical properties, predicted ADME, structural alerts, mutagenicity, and predicted PARP1 activity (Appendix~\ref{app:filters}). These filters remove molecules that fail basic selection criteria before more detailed evaluation. The surviving molecules are ranked using a weighted geometric mean of predicted PARP1 potency, toxicity, ADME, physicochemical properties, drug-likeness, synthetic accessibility, and novelty. This represents a conventional multi-parameter optimization approach in which favorable properties contribute to an overall desirability score. Notably, hERG and PARP2 are already included as continuous toxicity predictions. Clearance and fraction unbound, however, are not explicit ranking terms, although both influence exposure in the subsequent PBPK simulations. This distinction allows us to examine whether a score that already considers potency and off-target activity can still overlook exposure-related risks. The scoring functions and weights are described in Appendix~\ref{app:composite}.
To avoid selecting many closely related molecules, we construct a diversity-aware shortlist based on Murcko scaffolds, which describe the core ring and linker structure of a molecule. We first select the highest-ranked candidate from each scaffold before allowing a scaffold to contribute another molecule. This gives 20 shortlisted candidates per pipeline, or 60 molecules in total, for exposure-aware evaluation.

\textbf{Predictive models and PBPK implementation.}
We use Chemprop directed message-passing neural networks trained on experimental bioactivity data to predict PARP1 potency and the two safety endpoints, hERG and PARP2. Their held-out root mean squared errors (RMSE) are 0.79, 0.72, and 0.89 pIC50 units, respectively (Appendix~\ref{app:chemprop}). These predictions provide the target and off-target potency thresholds used in our dose and safety calculations.
We obtain the physicochemical and pharmacokinetic inputs required for PBPK simulation from ADMET-AI \citep{swanson2024admetai}, including lipophilicity, solubility, intestinal permeability, plasma protein binding, and hepatic clearance. Oral exposure is simulated using OSPSuite/PK-Sim 12.2 \citep{ospsuite}. We use the same physiological model and virtual individual for every candidate so that differences in predicted exposure arise from molecule-specific inputs rather than differences in the simulated individual (Appendix~\ref{app:pbpk-model}).
Finally, our safety classification accounts for uncertainty in both predicted off-target potency and simulated exposure. We use held-out Chemprop prediction errors and endpoint-specific PBPK errors reported by \citet{geci2024systematic}, as described in Appendix~\ref{app:pbpk-error}.

\section{Results}
\label{sec:results}

We evaluate whether shortlisted molecules achieve sufficient target exposure, remain safe at the required dose, and whether exposure-aware evaluation changes which candidates are preferred.

\begin{figure}[!t]
    \centering
    \includegraphics[width=0.9\linewidth]{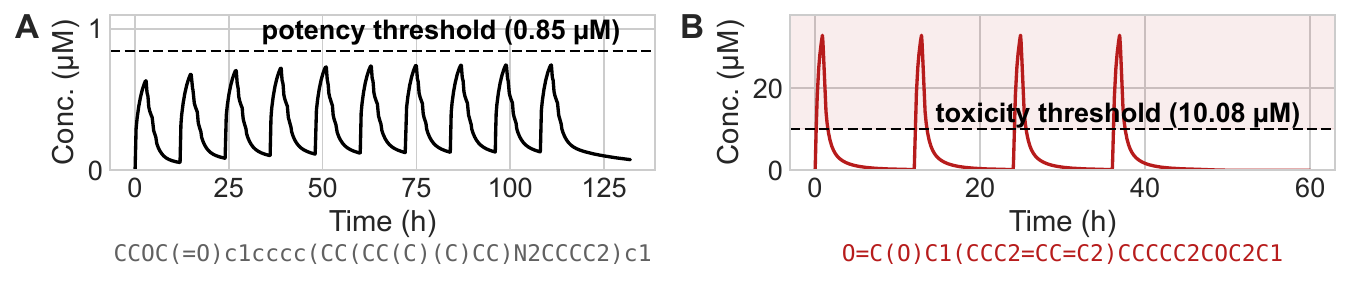}
    \vspace{-0.2cm}
    \caption{Concentration--time profiles after repeated dosing, illustrating two limitations revealed by PBPK simulation: (A) a compound that fails to reach its PARP1 potency threshold even at an extreme test dose; (B) a compound whose exposure at its MED exceeds the predicted hERG threshold. Thresholds are Chemprop-predicted IC50 values.}
    \label{fig:outcome-examples}
\end{figure}

\textbf{Can selected molecules achieve sufficient exposure?}
Seven of the 60 shortlisted molecules fail to reach the required target exposure under our dosing constraints: 5/20 from MOOD and 2/20 from Mol2Mol (Table~\ref{tab:outcome-summary}). The example in Figure~\ref{fig:outcome-examples} (left) fails to reach its PARP1 potency threshold even at an extreme test dose. These failures are not apparent from individual property predictions, as all seven molecules passed the same selection criteria as the others. Only when these properties are combined through PBPK simulation does their insufficient exposure become apparent (Appendix~\ref{app:nomed-gap}). The remaining 53 molecules can achieve sufficient exposure, but their required doses vary substantially within and across pipelines (Appendix~\ref{app:tables}, Figure~\ref{fig:med-by-regimen}). Although predicted PARP1 potency is associated with MED, molecules with similar potency can require very different doses because of differences in absorption, clearance, and protein binding.

\begin{wraptable}[14]{r}{0.5\textwidth}
    \centering
    \footnotesize
    \vspace{-1.1cm}
    \caption{Exposure and safety outcomes of generated candidates. \emph{Not sufficient}: no achievable MED. \emph{Risky}, \emph{Uncertain}, and \emph{Safe} correspond to the RED, ORANGE, and GREEN categories.}
    \label{tab:outcome-summary}
    \begin{tabular}{lccc}
        \toprule
         & MOOD & \shortstack{REINVENT4-\\MPO} & Mol2Mol \\
        \midrule
        Generated & 10{,}000 & 10{,}000 & 1{,}000 \\
        Post-filtering & 137 & 1{,}979 & 109 \\
        Shortlisted & 20 & 20 & 20 \\
        \midrule
        \multicolumn{4}{l}{\textit{Exposure} (of the 20 shortlisted)} \\
        Not sufficient & \textbf{5} & 0 & \textbf{2} \\
        \midrule
        \multicolumn{4}{l}{\textit{Safety} (of those with a dose)} \\
        Risky & \textcolor{red}{\textbf{7}} & \textcolor{red}{\textbf{2}} & \textcolor{red}{\textbf{6}} \\
        Uncertain & \textcolor{orange}{8} & \textcolor{orange}{17} & \textcolor{orange}{12} \\
        Safe & \textcolor{green!55!black}{0} & \textcolor{green!55!black}{1} & \textcolor{green!55!black}{0} \\
        \bottomrule
    \end{tabular}
\end{wraptable}
\textbf{Are selected molecules safe at the required exposure?}
Of the 53 candidates with an achievable MED, 15 (28\%) are classified as RED because their central predicted exposure exceeds at least one safety threshold: 7 from MOOD, 2 from REINVENT4-MPO, and 6 from Mol2Mol (Table~\ref{tab:outcome-summary}). Another 37 (70\%) are ORANGE, with apparently safe central estimates but overlapping uncertainty intervals. Only one candidate is GREEN for both endpoints.

The example in Figure~\ref{fig:outcome-examples} (right) has a low static hERG risk score of $P(\mathrm{active})=0.08$, yet its exposure at the MED exceeds the central predicted hERG IC50. This illustrates why static toxicity predictions alone can be misleading. Across all 53 candidates, predicted hERG risk is essentially uncorrelated with the exposure-adjusted hERG margin, whereas predicted PARP2 risk correlates much more strongly with its corresponding margin. Appendix~\ref{app:tables} provides the correlations and examines whether the raw off-target potency predictions reveal these risks.

\textbf{Does static ranking select the right molecules?}
We next examine whether better candidates were present in the generated pools but overlooked by static ranking. An unselected molecule dominates a shortlisted candidate if it requires a lower MED and has larger safety margins for both hERG and PARP2. To identify such alternatives efficiently, we first screen unselected molecules at a fixed dose, then apply the full MED procedure to promising candidates (Appendix~\ref{app:scoring}).
Among the 53 shortlisted molecules with an achievable MED, 45 (85\%) are dominated by at least one unselected candidate. These alternatives come from relatively small sets of confirmed dominators: 5 for MOOD, 108 for REINVENT4-MPO, and 12 for Mol2Mol. Thus, the generated pools already contain molecules with better predicted dose and safety profiles that the static ranking fails to select.

\section{Discussion and Limitations}
\label{sec:discussion}

Our results show that static molecular scores can select candidates with unfavorable dose requirements or exposure-related safety risks, even when their individual property predictions appear acceptable. More importantly, molecules with better predicted dose and safety profiles were often already present in the generated pools but overlooked during selection. This suggests that improving molecular generation alone may have limited benefit if candidate ranking does not account for how potency and pharmacokinetics jointly determine exposure at the dose needed for treatment.

Full PBPK simulation is too expensive to apply indiscriminately to large generated libraries, but static predictors remain useful for initial screening. A practical workflow could use static filters to narrow the candidate pool, exposure-aware estimates to guide ranking, and detailed PBPK simulations for final evaluation. Simply adding clearance or protein binding to a static score may not be sufficient, since their effects depend on other molecular properties and the required dose. This is consistent with previous work where adding pharmacokinetic objectives to molecular generation did not necessarily improve simulated therapeutic outcomes \citep{maagdenberg2025pkqsp}. Fast surrogate models trained on PBPK-derived dose requirements or safety margins could make exposure-aware selection more scalable, building on recent work in dose prediction and high-throughput PBPK modeling \citep{vanrompaey2024dose,geci2024systematic,naga2022htpbpk}.

These findings remain model-based rather than experimentally validated. Predicted potency thresholds and PBPK inputs are uncertain, and errors propagate into the estimated dose and safety margins. In particular, MED is calculated without propagating uncertainty in PARP1 potency (Appendix~\ref{app:med}). Our PBPK simulations also omit population variability and rely on global rather than compound-specific uncertainty estimates, which may be less reliable for novel molecules outside the training distribution. The RED, ORANGE, and GREEN categories therefore indicate predicted risk, not experimentally established safety or toxicity. Moreover, the magnitude of the ranking disagreement depends on the static scoring function and should not be interpreted as a universal failure rate. Our analysis is limited to PARP1 and three generative pipelines examined as independent case studies rather than a controlled comparison. Nevertheless, the results demonstrate that static selection can differ substantially from exposure-aware evaluation. Future work should examine whether this disagreement persists across other targets, scoring strategies, and experimental measurements.

\begin{ack}
This work was supported by a fellowship within the IFI programme of the German
Academic Exchange Service (DAAD), funded by the Federal Ministry of Education
and Research (BMBF). The authors declare no competing interests.
\end{ack}

\bibliographystyle{plainnat}

\clearpage
\appendix

\section{Minimum Effective Dose Search}
\label{app:med}

We determine the minimum effective dose (MED) separately for each molecule. The goal is to find the oral dose at which the predicted free steady-state trough concentration reaches the PARP1 potency threshold. We begin with a single PK-Sim simulation at a reference dose of 70~mg and use linear scaling to estimate an initial dose. Each subsequent iteration runs a new oral PBPK simulation, estimates the terminal half-life, and constructs the concentration--time profile after repeated dosing. We then compare the free steady-state trough concentration, $C_{ss,\min,\mathrm{free}}$, with the predicted PARP1 IC50. We consider once-daily (QD) and twice-daily (BID) dosing, since the dosing interval affects how much the drug concentration declines between administrations. A molecule with a short half-life may fall below the target concentration between doses, even if its peak concentration is sufficiently high. We select the regimen using the half-life estimated in the first iteration and keep it fixed throughout the search. This prevents molecules near the QD/BID boundary from repeatedly switching regimens, which could cause the search to oscillate rather than converge.

To update the dose, we use a log--log secant step toward a concentration of $1.05\times IC50_{\mathrm{target}}$. The small offset above the threshold provides a buffer against numerical fluctuations. Each update is limited to a fourfold change, with doses constrained to 0.1--1{,}000{,}000~mg and a maximum of 30 iterations. If the secant search oscillates or revisits a previous dose, we switch to bisection between the closest doses that bracket the target. The search terminates when the predicted exposure is within 3\% of the target, progress stalls, or an iteration or dose limit is reached. Candidates that never reach the target concentration within these constraints are reported as having no achievable MED. The upper limit of 1{,}000{,}000~mg is far beyond any clinically practical dose. It is intended to distinguish candidates whose predicted exposure cannot reach the target within our model and search range from those that merely require an unusually large dose. Importantly, finding an MED does not imply that the corresponding dose is clinically feasible.

The search uses the central predicted PARP1 pIC50 without propagating uncertainty in that prediction. Its held-out RMSE of 0.79 pIC50 units corresponds to approximately a sixfold difference in concentration. Because the estimated MED also determines the exposure used for both safety endpoints (Appendix~\ref{app:pbpk-error}), uncertainty in target potency affects both the reported dose and safety margins. This additional source of uncertainty is not quantified in our analysis.

\section{Safety Uncertainty and PBPK Benchmark}
\label{app:pbpk-error}

Our safety assessment accounts for uncertainty in both predicted off-target potency and simulated drug exposure. This matters because a candidate whose central exposure estimate falls below a safety threshold may still exceed it when prediction errors are considered. We use endpoint-specific PBPK errors because hERG and PARP2 are evaluated using different parts of the concentration--time profile. For hERG, which is assessed using peak concentration ($C_{\max}$), we use the compound-level oral $C_{\max}$ error reported by \citet{geci2024systematic}, with a median $\log_2$ MAE of 1.13. For PARP2, which is evaluated using steady-state trough exposure, we use their larger point-by-point concentration--time-profile error (median $\log_2$ MAE of 1.78). This latter metric includes late-phase measurements and provides a more conservative approximation of trough uncertainty, which was not reported separately. We use the benchmark from \citet{geci2024systematic} because it evaluates 161 compounds, providing a broader empirical basis for estimating PBPK prediction error than a small benchmark constructed from our generated candidates. However, the benchmark uses different in-silico ADME predictors from ADMET-AI, which supplies the inputs to our own PBPK simulations. The reported errors may therefore not fully capture the uncertainty of our specific implementation.

For off-target potency, we use the held-out RMSE of the corresponding Chemprop model: 0.72 pIC50 units for hERG and 0.89 for PARP2 (Appendix~\ref{app:chemprop}). These errors are considered alongside PBPK exposure uncertainty when assigning safety categories. A candidate is classified as \emph{RED} if its central simulated exposure exceeds the central predicted off-target IC50. It is \emph{ORANGE} if the central exposure is below the threshold but the uncertainty intervals overlap, and \emph{GREEN} if the intervals remain separated. A candidate's overall classification reflects its assessment across both safety endpoints. These categories describe risk under our predictive models, not experimentally established toxicity or safety. In particular, GREEN means that the modeled uncertainty intervals do not indicate a safety concern for the evaluated endpoints. It does not establish that a molecule is safe in vivo or free from other toxicities. The PBPK implementation is described in Appendix~\ref{app:pbpk-model}.

\section{Generative Pipeline Implementation}
\label{app:generation}

All three generative pipelines use publicly available implementations. MOOD and REINVENT4-MPO generate molecules de novo, whereas Mol2Mol modifies existing molecules through seed-based generation. The pipelines differ in how they explore chemical space and which properties influence generation. \emph{MOOD} \citep{lee2023mood} uses the released GDSS diffusion backbone pretrained on ZINC250k. Generation is guided by the publicly available PARP1 docking-score regressor (\texttt{prop\_parp1}), which predicts docking scores from noised molecular graphs rather than performing molecular docking at every diffusion step. We use the repository's default out-of-distribution control strength, \texttt{ood=0.04}. MOOD does not explicitly optimize pharmacokinetic or ADME properties during generation.

\emph{REINVENT4-MPO} uses the public REINVENT4 framework \citep{loeffler2024reinvent4}. Its recurrent neural network is fine-tuned through reinforcement learning using a multi-parameter reward. PARP1 potency is estimated by a Random Forest surrogate trained on ChEMBL bioactivity data using Morgan fingerprints. Fine-tuning runs for 200 steps with a batch size of 64, corresponding to approximately 12{,}800 scoring calls. Unlike MOOD, the optimization objective includes physicochemical and ADME-related properties alongside predicted potency. \emph{Mol2Mol} uses REINVENT4's molecule-to-molecule generation mode rather than a separate model implementation. We initialize generation from 20 approved PARP1 inhibitors and produce 50 analogues per seed, giving 1{,}000 candidates. This approach explores chemical neighborhoods around existing molecules rather than generating candidates entirely de novo. We therefore sample fewer molecules from Mol2Mol than from MOOD and REINVENT4-MPO, which each generate 10{,}000 candidates.

\paragraph{Why the pipelines are not directly comparable.}
\label{app:confound}
The three pipelines differ in architecture, training data, starting chemical space, and optimization objective. We use them as independent case studies to examine whether the same exposure-related selection problems arise in different generative settings, not to determine which generative method performs best. For example, the relationship between predicted potency and MED differs across pipelines, but this cannot be attributed solely to whether ADME properties were included during optimization. Establishing such a relationship would require a controlled experiment in which architecture, training data, and other conditions are held fixed while the optimization objective changes. We therefore interpret differences within the context of each pipeline and avoid drawing causal conclusions about their relative pharmacokinetic performance.

\section{Filtering Criteria}
\label{app:filters}

All generated molecules undergo the same hard filtering procedure before static ranking. The purpose is to remove candidates with unfavorable predicted properties while retaining a pool for multi-parameter selection. A molecule is retained only if it satisfies all of the following criteria: QED (Quantitative Estimate of Drug-likeness) $\geq 0.5$; synthetic accessibility score $\leq 6.0$; TPSA (topological polar surface area) between 20 and 140~\AA$^2$; molecular weight $\leq 500$~Da; predicted logP between 1.0 and 5.0; predicted logS $\geq -6.0$; Caco-2 permeability (logP$_{\mathrm{app}}$) $\geq -5.15$; hydrogen-bond donors $\leq 5$; hydrogen-bond acceptors $\leq 10$; plasma protein binding $\leq 99\%$; Ames mutagenicity risk $\leq 0.5$; no PAINS or Brenk structural alerts; and predicted PARP1 pIC50 $\geq 6.0$.

These thresholds serve different purposes. For example, the plasma protein binding cutoff removes molecules predicted to be more than 99\% bound, rather than treating lower binding as universally preferable. Likewise, Ames mutagenicity is used as a hard exclusion because it represents a distinct toxicity concern. In contrast, hERG and PARP2 are included as continuous, uncertainty-aware desirability terms in the subsequent ranking rather than as hard filters, consistent with multi-parameter optimization practice \citep{segall2012mpo}. This avoids discarding candidates solely because of one uncertain off-target prediction. It also allows us to examine whether these static toxicity estimates agree with the exposure-adjusted safety margins obtained later. The filters retain 137 of 10{,}000 MOOD candidates (1.4\%), 1{,}979 of 10{,}000 REINVENT4-MPO candidates (19.8\%), and 109 of 1{,}000 Mol2Mol candidates (10.9\%). These surviving molecules form the pools used for static ranking.

\section{Static Candidate Ranking}
\label{app:composite}

After filtering, candidates are ranked using a weighted geometric mean (WGM) of static molecular desirability terms:
\[
\mathrm{WGM}
=
\prod_i d_i^{w_i/\sum_j w_j},
\]
where $d_i \in [0,1]$ represents the desirability of property $i$ and $w_i$ its weight. The geometric mean penalizes candidates with poor values for individual properties more strongly than an arithmetic mean. A high score on one property therefore cannot fully compensate for a very low score on another. Predicted PARP1 potency receives the highest weight, 2.0, and is represented by the Chemprop ensemble probability that pIC50 $\geq 6$. QED, synthetic accessibility, hERG risk, PARP2 selectivity, CYP3A4 risk, aqueous solubility, and Caco-2 permeability each receive weight 1.0. CYP3A4 risk serves as a proxy for metabolic drug--drug interaction liability. Molecular weight, logP, TPSA, and intellectual property (IP) novelty each receive weight 0.5.

For hERG and PARP2, desirability is defined as $1-P(\mathrm{active})$, where
\[
P(\mathrm{active})
=
\Phi\!\left(
\frac{\mathrm{pIC50}-6}{\mathrm{RMSE}}
\right).
\]
Here, $\Phi$ denotes the standard normal cumulative distribution function, and RMSE is the held-out error of the corresponding Chemprop model (Appendix~\ref{app:chemprop}). This transformation maps predicted potency to a continuous activity score rather than assigning risk through a single hard threshold. Solubility and permeability receive increasing desirability above their filtering thresholds. Molecular weight, logP, and TPSA use bell-shaped functions centered at 350~Da, 2.5, and 75~\AA$^2$, respectively, favoring intermediate values. Novelty is based on Tanimoto similarity to approved drugs, with desirability decreasing as similarity increases from 0.7 to 1.0. We rank candidates by WGM while imposing a diversity constraint: the highest-ranked molecule from each Murcko scaffold is selected before any scaffold contributes a second candidate. Murcko scaffolds represent the core ring and linker structures of molecules, so this procedure reduces the chance that the shortlist consists mainly of close structural analogues. We select 20 molecules per pipeline.

Clearance and fraction unbound are not explicit terms in the static ranking, although both influence PBPK exposure. Neither has a universally favorable direction independent of the other molecular properties. For example, low clearance may increase exposure but can also increase concentrations associated with off-target activity. Their omission alone does not explain the disagreement with exposure-aware safety evaluation. RED candidates do not consistently have extreme clearance or fraction-unbound values. Instead, the disagreement arises from how pharmacokinetic properties combine with target and off-target potency to determine exposure and safety at the required dose.

\section{Potency and Safety Prediction Models}
\label{app:chemprop}

We predict PARP1, hERG, and PARP2 bioactivity using Chemprop directed message-passing neural networks trained on experimental bioactivity data. These models provide the potency thresholds used for target engagement and exposure-adjusted safety assessment. Training records are obtained from ChEMBL, restricting entries to confidence scores $\geq 8$ on ChEMBL's 0--9 assay-to-target mapping reliability scale and exact activity relations. Duplicate SMILES are collapsed to the median pIC50 to avoid treating multiple measurements of the same molecular structure as independent compounds.

The PARP1 model uses 3{,}463 compounds for target CHEMBL3105, while the hERG model uses 8{,}256 compounds for CHEMBL240. The initial PARP2 dataset contains 719 compounds for CHEMBL5366 and is expanded to 1{,}189 using additional ChEMBL Ki/Kd and BindingDB records associated with UniProt Q9UGN5. Each model uses an 80/10/10 scaffold-balanced train/validation/test split. Held-out root mean squared error (RMSE) is 0.79 pIC50 units for PARP1, 0.72 for hERG, and 0.89 for the expanded PARP2 dataset. Reported PARP1 performance pools predictions from two replicates, hERG from one, and PARP2 from three. Because pIC50 is logarithmic, even a moderate prediction error can correspond to a substantial difference in the concentration needed for inhibition. We account for the reported off-target prediction errors in the safety classification, while the PARP1 MED search uses the central potency estimate.

\section{PBPK Model Construction}
\label{app:pbpk-model}

All candidates are evaluated using the same OSPSuite/PK-Sim physiological model and virtual individual. This keeps the physiological assumptions fixed, so differences between simulated concentration profiles arise from molecule-specific inputs and dosing rather than changes in the simulated individual.

\paragraph{Physiological model.}
We use PK-Sim's standard whole-body physiologically based model, which represents major organs and tissues as compartments connected through systemic blood flow. The simulations use a single virtual individual (70~kg, 175~cm, male, age 43, European/ICRP population). Organ volumes and blood flows are taken from PK-Sim's physiology database. Tissue-to-plasma partitioning uses PK-Sim's Standard method for all compartments except the interstitial compartment, which uses the Schmitt method. We do not simulate population variability, so the predicted exposure profiles represent one modeled physiological setting rather than the range of exposures expected across patients.

\paragraph{Molecule-specific inputs.}
We use ADMET-AI \citep{swanson2024admetai} to predict lipophilicity, solubility, intestinal permeability, plasma protein binding, and hepatic clearance for each molecule. These properties determine how the PBPK model represents absorption, distribution, and elimination, and thereby influence the resulting concentration--time profile. All candidates are evaluated following oral administration, using the same route for the trial-dose simulations during the MED search and the final repeated-dose steady-state evaluation.

\paragraph{Clearance.}
Hepatic elimination is represented using PK-Sim's well-stirred liver model, with specific hepatic clearance obtained from ADMET-AI predictions. Renal elimination follows PK-Sim's default physiological glomerular filtration. Together, these processes contribute to the predicted decline in drug concentration between doses and therefore affect the dose required to maintain the target trough exposure.

\section{Supplementary Exposure and Safety Results}
\label{app:tables}

\paragraph{MED across and within pipelines.}
\begin{figure}[h]
    \centering
    \includegraphics[width=\linewidth]{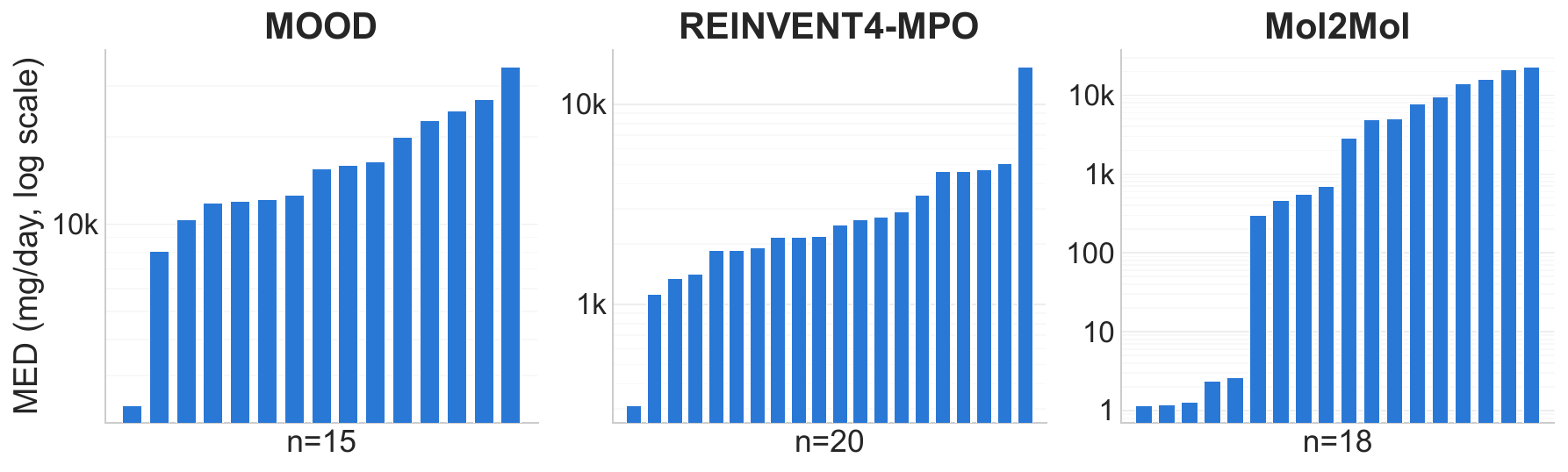}
    \caption{Minimum effective dose (MED, mg/day, logarithmic scale) for the 53 shortlisted candidates with an achievable dose, ranked within each case study. MOOD has 15 candidates, REINVENT4-MPO has 20, and Mol2Mol has 18; the remaining seven candidates have no achievable MED. Required doses vary by several orders of magnitude within and across pipelines.}
    \label{fig:med-by-regimen}
\end{figure}

Median MED is 1{,}785~mg/day for Mol2Mol (range 1.1--22{,}658), 2{,}340~mg/day for REINVENT4-MPO (range 310--15{,}326), and 15{,}487~mg/day for MOOD (range 2{,}357--34{,}793). Figure~\ref{fig:med-by-regimen} shows substantial variation within each pipeline, indicating that molecules selected using similar static criteria can require very different doses. These differences should not be interpreted as a controlled comparison of the generators, which differ in several aspects of their design and training.

\paragraph{Predicted potency and static risk versus PBPK-derived outcomes.}
We examine how well the static Chemprop scores agree with their corresponding exposure-aware outcomes. For the 53 shortlisted candidates with an achievable MED, Table~\ref{tab:pooled-corr} reports pooled Spearman rank correlations between PARP1 potency and MED, hERG risk and its exposure-adjusted safety margin, and PARP2 risk and its corresponding margin. PARP1 potency is strongly associated with MED ($\rho=-0.81$), as more potent molecules generally require lower doses. However, potency alone does not fully determine dose, since absorption, clearance, and protein binding also affect the concentration profile. The two safety endpoints show different patterns: predicted hERG risk is essentially uncorrelated with its exposure-adjusted margin ($\rho=-0.03$), whereas PARP2 risk shows a strong negative correlation ($\rho=-0.70$). Thus, the usefulness of static off-target predictions depends on how well they reflect risk at the exposure required for target engagement.

\begin{table}[h]
    \caption{Spearman correlations between static Chemprop scores and PBPK-derived outcomes, pooled across the 53 shortlisted candidates with an achievable MED.}
    \label{tab:pooled-corr}
    \centering
    \begin{tabular}{lcc}
        \toprule
        Comparison & $\rho$ & $p$ \\
        \midrule
        PARP1 potency vs.\ MED         & $-0.81$ & $<10^{-8}$ \\
        hERG risk vs.\ safety margin   & $-0.03$ & $0.85$ \\
        PARP2 risk vs.\ safety margin  & $-0.70$ & $<10^{-8}$ \\
        \bottomrule
    \end{tabular}
\end{table}

\paragraph{Would a simple risk threshold work instead?}
One alternative to PBPK-based safety evaluation would be to apply a fixed off-target potency threshold. For example, using a hERG IC50 cutoff of 10~$\mu$M, as in the TDC hERG benchmark \citep{wang2016herg}, would classify 6 of the 13 hERG-RED candidates as apparently safe. At the same time, it would flag 7 of the 8 hERG-GREEN candidates as risky. A fixed threshold therefore misses some candidates with high exposure-related risk while excluding others with more favorable predicted safety margins. A threshold based on predicted activity probability, such as $P(\mathrm{active})\geq0.5$, performs better for PARP2. It would identify all eight PARP2-RED candidates but still classify 20 of the 41 PARP2-ORANGE candidates (49\%) as apparently safe. These results illustrate why a single potency threshold cannot fully replace evaluation at the required exposure, particularly when predictions are uncertain.

\paragraph{Individual parameters of candidates with shortcomings.}
\label{app:nomed-gap}

We next examine whether the seven candidates without an achievable MED and the 15 candidates classified as RED could have been identified from individual molecular properties alone. Tables~\ref{tab:nomed-pk}--\ref{tab:red-pk} report their predicted physicochemical and pharmacokinetic parameters. The seven candidates without an achievable MED tend to have lower unbound fractions and higher clearance than candidates with an achievable dose. Their median fraction unbound is 0.12, compared with 0.28 for the remaining candidates (two-sided Mann--Whitney $p=0.0006$). Median specific clearance is 19.5 versus 9.31 ($p=0.016$). Yet neither property consistently falls outside a plausible drug-like range. Instead, their combined effects, together with potency and absorption, limit the free concentration that can be maintained between doses. An individual prediction may therefore satisfy its acceptance threshold even when the combination of properties produces insufficient exposure.

\begin{table}[h]
    \caption{Predicted physicochemical and pharmacokinetic parameters for the seven candidates without an achievable MED. Values are obtained from ADMET-AI.}
    \label{tab:nomed-pk}
    \centering
    \footnotesize
    \begin{tabular}{lccccccc}
        \toprule
        Molecule & Model & logP & logS & Perm. & $f_u$ & CL & MW (Da) \\
        \midrule
        MOOD\_P1\_005 & MOOD    & 3.54 & $-3.89$ & $-4.46$ & 0.124 & 22.6 & 286 \\
        MOOD\_P1\_006 & MOOD    & 1.33 & $-3.98$ & $-4.44$ & 0.064 & 8.1  & 302 \\
        MOOD\_P1\_010 & MOOD    & 4.35 & $-5.06$ & $-4.46$ & 0.037 & 22.3 & 342 \\
        MOOD\_P1\_016 & MOOD    & 3.09 & $-4.01$ & $-4.89$ & 0.120 & 16.5 & 332 \\
        MOOD\_P1\_018 & MOOD    & 2.95 & $-4.41$ & $-4.67$ & 0.208 & 21.0 & 332 \\
        Mol2Mol\_009  & Mol2Mol & 3.33 & $-4.86$ & $-4.59$ & 0.044 & 2.9  & 293 \\
        Mol2Mol\_011  & Mol2Mol & 2.27 & $-4.44$ & $-5.03$ & 0.134 & 19.5 & 384 \\
        \bottomrule
    \end{tabular}
\end{table}

Tables~\ref{tab:red-risk} and \ref{tab:red-pk} provide the MED, predicted off-target potency, and pharmacokinetic parameters for the 15 RED candidates. The raw potency values help explain why the two safety endpoints behave differently. The eight PARP2-RED candidates have a median PARP2 IC50 of 0.26~$\mu$M (range 0.12--0.47), indicating relatively strong predicted PARP2 inhibition. In contrast, the 13 hERG-RED candidates have a median hERG IC50 of 8.32~$\mu$M (range 3.75--13.81). Six of these 13 candidates exceed the 10~$\mu$M hERG threshold and would therefore appear less concerning under that static criterion. For these candidates, exposure-adjusted risk is not explained by unusually strong predicted hERG inhibition alone. Rather, the concentration reached at the dose required for PARP1 engagement is sufficiently high relative to the predicted hERG threshold. A static potency cutoff cannot capture this dependence on exposure.

\begin{figure}[h]
    \centering
    \includegraphics[width=\linewidth]{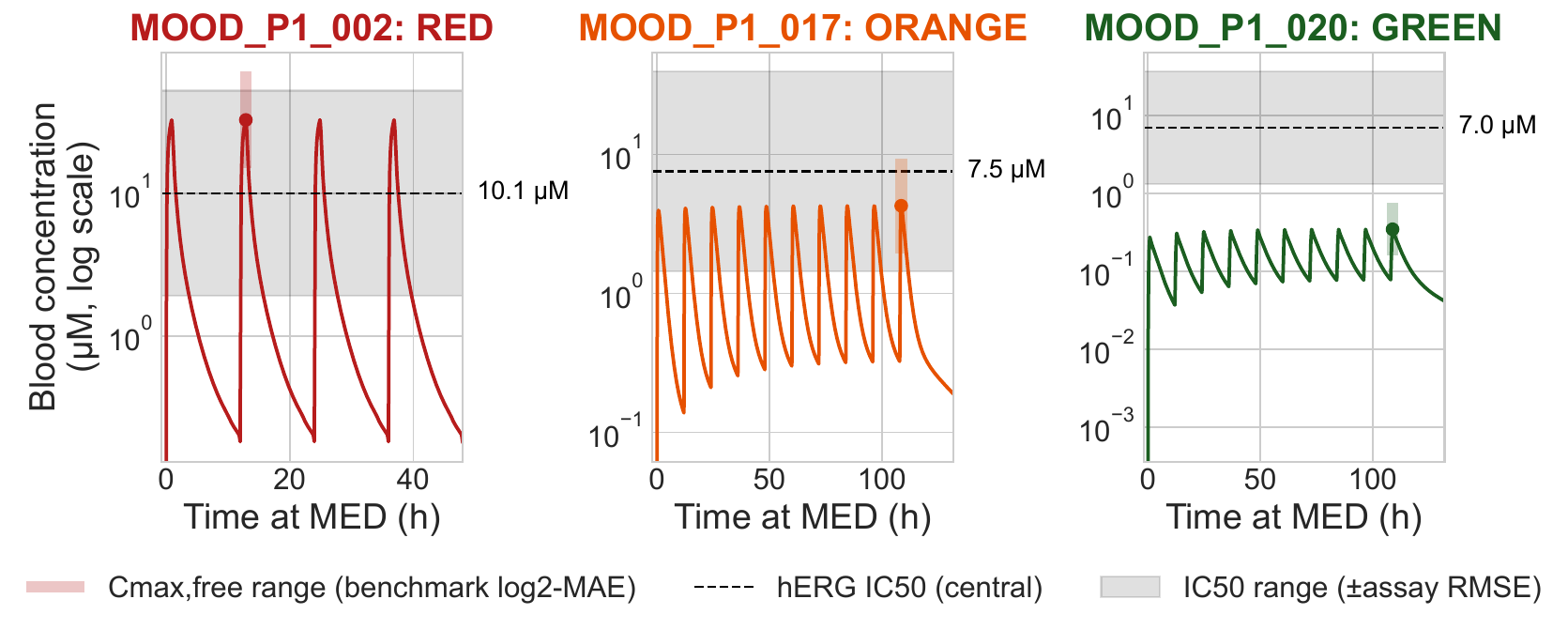}
    \caption{Simulated blood concentrations at each candidate's MED for three MOOD compounds representing the hERG RED, ORANGE, and GREEN categories. All have similar predicted hERG IC50 values (7.0--10.1~$\mu$M). Shaded horizontal bands show IC50 uncertainty (\textpm~assay RMSE), while vertical bars at the peaks indicate uncertainty in free $C_{\max}$ (benchmark $\log_2$ MAE). A 10~$\mu$M hERG cutoff would classify the RED example (IC50 10.1~$\mu$M) as apparently safe and the GREEN example (IC50 7.0~$\mu$M) as risky. Exposure-aware evaluation reverses this interpretation: RED exceeds the central IC50, ORANGE has overlapping uncertainty intervals, and GREEN remains below the threshold even after accounting for uncertainty.}
    \label{fig:example-safety-curve}
\end{figure}

\paragraph{Raw hERG IC50 is not monotonic across risk tiers.}
The difference between hERG and PARP2 also appears across their endpoint-specific risk categories. Median PARP2 IC50 decreases from GREEN (3.00~$\mu$M) to ORANGE (0.98~$\mu$M) to RED (0.26~$\mu$M), consistent with stronger predicted inhibition corresponding to greater risk. For hERG, the relationship is not monotonic: median IC50 is 7.09~$\mu$M for GREEN, 10.62~$\mu$M for ORANGE, and 8.32~$\mu$M for RED. Moreover, 6 of the 8 hERG-GREEN candidates have a lower, apparently more potent hERG IC50 than the median hERG-RED candidate. This illustrates why raw hERG potency alone does not reliably distinguish exposure-adjusted risk categories in our candidate set. Figure~\ref{fig:example-safety-curve} illustrates this difference using three MOOD candidates with similar predicted hERG potency but different exposure-aware classifications.

\begin{table}[h]
    \caption{The 15 candidates classified as RED for hERG and/or PARP2, showing MED and predicted off-target IC50 values. Bold indicates the endpoint responsible for the RED classification.}
    \label{tab:red-risk}
    \centering
    \footnotesize
    \begin{tabular}{llccc}
        \toprule
        Molecule & Model & MED (mg/day) & hERG IC50 ($\mu$M) & PARP2 IC50 ($\mu$M) \\
        \midrule
        MOOD\_P1\_007 & MOOD          & 12{,}565 & \textbf{8.29}  & \textbf{0.47} \\
        MOOD\_P1\_002 & MOOD          & 16{,}451 & \textbf{10.08} & 0.55          \\
        MOOD\_P1\_019 & MOOD          & 19{,}881 & \textbf{4.64}  & \textbf{0.28} \\
        MOOD\_P1\_015 & MOOD          & 22{,}842 & 7.38           & \textbf{0.46} \\
        MOOD\_P1\_011 & MOOD          & 24{,}542 & \textbf{10.21} & 0.55          \\
        MOOD\_P1\_008 & MOOD          & 26{,}869 & \textbf{5.28}  & \textbf{0.27} \\
        MOOD\_P1\_001 & MOOD          & 34{,}793 & \textbf{8.32}  & 1.05          \\
        R4MPO\_010    & REINVENT4-MPO & 5{,}041  & \textbf{11.18} & 2.24          \\
        R4MPO\_005    & REINVENT4-MPO & 15{,}326 & \textbf{13.81} & 2.40          \\
        Mol2Mol\_017  & Mol2Mol       & 298      & 4.04           & \textbf{0.21} \\
        Mol2Mol\_019  & Mol2Mol       & 4{,}861  & \textbf{5.37}  & 0.30          \\
        Mol2Mol\_018  & Mol2Mol       & 7{,}677  & \textbf{3.77}  & \textbf{0.25} \\
        Mol2Mol\_002  & Mol2Mol       & 14{,}188 & \textbf{12.91} & \textbf{0.15} \\
        Mol2Mol\_015  & Mol2Mol       & 15{,}913 & \textbf{10.74} & \textbf{0.12} \\
        Mol2Mol\_010  & Mol2Mol       & 21{,}118 & \textbf{3.75}  & 1.09          \\
        \bottomrule
    \end{tabular}
\end{table}

\begin{table}[h]
    \caption{ADMET-AI-predicted physicochemical and pharmacokinetic parameters for the 15 RED candidates.}
    \label{tab:red-pk}
    \centering
    \footnotesize
    \begin{tabular}{llccccc}
        \toprule
        Molecule & Model & logP & logS & Perm. & $f_u$ & CL \\
        \midrule
        MOOD\_P1\_007 & MOOD          & 1.53 & $-2.38$ & $-4.42$ & 0.490 & 4.0  \\
        MOOD\_P1\_002 & MOOD          & 1.02 & $-3.08$ & $-4.83$ & 0.150 & 6.6  \\
        MOOD\_P1\_019 & MOOD          & 2.47 & $-3.27$ & $-4.55$ & 0.231 & 9.0  \\
        MOOD\_P1\_015 & MOOD          & 3.31 & $-3.71$ & $-4.04$ & 0.221 & 15.7 \\
        MOOD\_P1\_011 & MOOD          & 1.84 & $-1.02$ & $-4.11$ & 0.500 & 18.5 \\
        MOOD\_P1\_008 & MOOD          & 2.81 & $-2.53$ & $-3.83$ & 0.351 & 15.1 \\
        MOOD\_P1\_001 & MOOD          & 2.94 & $-3.17$ & $-4.18$ & 0.265 & 17.9 \\
        R4MPO\_010    & REINVENT4-MPO & 1.46 & $-0.96$ & $-4.59$ & 0.557 & 15.3 \\
        R4MPO\_005    & REINVENT4-MPO & 1.36 & $-2.26$ & $-4.30$ & 0.335 & 9.0  \\
        Mol2Mol\_017  & Mol2Mol       & 2.20 & $-4.88$ & $-4.60$ & 0.022 & 0.6  \\
        Mol2Mol\_019  & Mol2Mol       & 1.59 & $-2.16$ & $-4.63$ & 0.496 & 4.9  \\
        Mol2Mol\_018  & Mol2Mol       & 2.30 & $-2.97$ & $-5.04$ & 0.153 & 7.9  \\
        Mol2Mol\_002  & Mol2Mol       & 1.09 & $-3.48$ & $-4.55$ & 0.243 & 14.2 \\
        Mol2Mol\_015  & Mol2Mol       & 1.85 & $-3.51$ & $-4.63$ & 0.040 & 7.2  \\
        Mol2Mol\_010  & Mol2Mol       & 2.20 & $-3.02$ & $-4.57$ & 0.265 & 10.0 \\
        \bottomrule
    \end{tabular}
\end{table}

\section{Static Ranking Versus Exposure-Aware Selection}
\label{app:scoring}

\paragraph{Searching the unselected pool.}
To assess whether static ranking overlooks more favorable candidates, we search the generated pools for molecules that outperform the shortlisted candidates on required dose and both safety margins. Running a full individualized MED search for every filtered molecule would require evaluating up to 1{,}979 candidates in a single pipeline. We therefore use a two-stage procedure to reduce computational cost. First, we evaluate unselected molecules at a fixed oral dose of 70~mg. We retain candidates classified as GREEN for both safety endpoints that appear capable of outperforming at least one shortlisted molecule on target engagement, hERG margin, and PARP2 margin. This screening step identifies 158 potential dominators: 18 from MOOD, 119 from REINVENT4-MPO, and 21 from Mol2Mol. The fixed-dose evaluation is only a preliminary screen because a molecule that looks promising at one dose may behave differently when dosed to maintain its own target exposure. We therefore run the full individualized MED search from Appendix~\ref{app:med} for all 158 candidates. Of these, 33 (21\%) have no achievable MED, leaving 5 confirmed dominators from MOOD, 108 from REINVENT4-MPO, and 12 from Mol2Mol.

We compare candidates only after determining their individualized MED. An unselected molecule dominates a shortlisted candidate if it has a lower MED and larger safety margins for both hERG and PARP2. This definition requires improvement on all three quantities simultaneously, rather than a trade-off between lower dose and greater safety. Under this criterion, 93\% of MOOD, 100\% of REINVENT4-MPO, and 61\% of Mol2Mol shortlisted candidates with an achievable MED are dominated. Overall, 45 of the 53 candidates (85\%) are outperformed by at least one unselected molecule. The alternatives are not simply repeated variants of compounds already represented in the shortlist: every confirmed dominator has a Murcko scaffold absent from its corresponding shortlist. The MOOD dominators occupy 3 distinct scaffolds among 5 molecules, REINVENT4-MPO occupies 103 among 108, and Mol2Mol occupies 11 among 12. Thus, the missed alternatives also represent structural diversity that the static shortlist does not capture.

\end{document}